\documentclass[
 reprint,
 amsmath,amssymb,
 aps,
]{revtex4-2}

\usepackage{graphicx}
\usepackage{dcolumn}
\usepackage{bm}
\usepackage{tikz}

\begin{document}

\preprint{APS/123-QED}

\title{Definition of vortex boundary using stagnation pressure}

\author{Marc Plasseraud}
\author{Krishnan Mahesh}
 \email{krmahesh@umich.edu}
\affiliation{
 Naval Architecture \& Marine Engineering, \\
 University of Michigan
}

\begin{abstract}
A novel method is proposed to identify vortex boundary and center of rotation based on tubular surfaces of constant stagnation pressure and minimum of the stagnation pressure gradient. The method is derived from Crocco's theorem, which ensures that the gradient of stagnation pressure is orthogonal to both the velocity and vorticity vectors. The method is Galilean invariant, requires little processing and is robust. It enables visualization of complex turbulent flows and provides a physically consistent definition of vortex boundaries for quantitative analyses. This vortex boundary is a material surface that is representative of the kinematics of the flow by construction, constitutes a vortex tube, ensures conservation of circulation in the inviscid limit and provides a unique relation to the conservation of momentum equations and vortex loads.
\end{abstract}


\maketitle

\paragraph*{Introduction}
 Vortices are a near universal and characteristic feature of fluid flows. Defining their boundary and centerline is crucial for understanding flow behavior, loads and more fundamentally, energy transport in turbulence. While they are visually easy to identify, there is no consensus on the mathematical definition of what constitutes a vortex, and what \citet{lugt1979} aptly called the `dilemna of defining a vortex' is still a topic of active research. \citet{robinson1991} groups vortices into a more general category of `coherent motions' which is the foundation of turbulent structures. This definition perhaps hints at the difficulty encountered in trying to use a metric that is local and instantaneous to describe a phenomenon that is fundamentally non-local (spatial coherence) and transient (temporal coherence) while being usable across a wide variety of flows. Several methods have been proposed and successfully applied \citep{holmen2012, epps2017}; the difficulty generally resides in finding a metric that can accommodate a large variety of vortex types while distinguishing between shearing and vortical vorticity and being invariant \cite{haller2005}. In their review of vortex identification, \citet{epps2017} formalizes a list of requirements for methods, which includes among others: be objective (invariant to translation and rotation of the coordinate system); avoid use of `an arbitrary threshold';  identify coherent motion; guarantee non-zero vorticity; be insensitive to shear vorticity; be applicable to compressible flows and heterogeneous fluids; be robust to experimental noise. In addition to these criteria, secondary properties are desirable: a method should be computationally inexpensive and favor the use of local variables, and not depend on higher order quantities that can be slow to converge thus require long sampling times. Many methods exist that have verified some part of these requirements \cite{holmen2012}, \cite{epps2017} although none validate all of them. Some of the most commonly used methods are :
\begin{itemize}
    \item pressure \cite{robinson1991}: vortices can be identified as minima of pressure. This method has the advantage of simplicity and being invariant, however the presence of a minimum of pressure does not guarantee the existence of a vortex. Worse, the presence of a vortex does not guarantee a minimum of pressure such as the case of vortices with strong axial velocity (see paragraph `Case II - 6:1 prolate spheroid'). Instead of using pressure directly, \citet{li2020} used the gradient of pressure normalized by its Laplacian as a distance field to the vortex core. This approach recovers the radius of a Rankine vortex and shows promising results for the cases demonstrated in their study such as homogeneous isotropic turbulence however, as the authors note, it is limited to cases where the pressure gradient originates from the vortex;
    \item vorticity magnitude or axial vorticity \cite{saffman1995}: \citet{saffman1995} defines vortices as a connected region of vorticity. In accordance with this definition, vortices can be identified through a maximum of axial vorticity although this requires a-priori knowledge of the axis of vorticity; or through the maximum of enstrophy \cite{comte1998}. This method is simple and the presence of a vortex guarantees the existence of a local extrema of vorticity, however, it does not distinguish between shear layer and vortices;
    \item The Q criterion method, proposed by \citet{hunt1988} and refined by \citet{dubief2000} using large-eddy simulation and direct numerical simulation, identifies vortices as regions where the Q value is positive with:
    \[
    Q = ||\Omega||_2^2 - ||S||_2^2
    \]
    where $S$ is the strain rate tensor and $\Omega$ is the rotation tensor. Positive values are indicative of regions with strong streamline curvature; 
    \item \citet{jeong1995} proposed another velocity gradient tensor based method based on the second eigenvalue of the matrix $\Omega^2 + S^2$. It uses the fact that a negative second eigenvalue is indicative of a rotation-dominated area of the flow, to identify coherent rotating motion. Both the Q criterion and $\lambda_2$ methods are widely used and identify vortices in strong vorticity regions and discriminate efficiently between shear layer and coherent motion, however they can generate false positives (corner flow for example) and may not align as well with a more intuitive streamline definitions (see paragraph `cavity flow'). Their limitation comes from the exclusive use of kinematic variables and it will be shown that they fail in the case of a cavity flow where the streamlines are curved outside a vortex;
    \item Lagrangian based \cite{robinson1989}: these methods define vortices in terms of particle trajectory such as closed streamlines. While they give results that are close to the qualitative observation of vortices, they are harder to use since they rely on non-local and time-varying variables. In addition, it can difficult to assess whether streamlines are closed, especially for three-dimensional vortices. Furthermore, using velocity vectors is not an objective metric. It may give a different result if a constant velocity component is added to the field \cite{lugt1979}.
\end{itemize}

Despite the multitude of methods available, none offers both a robust way of visualizing vortices in complex flow and a way to identify vortex topology in accordance with the kinematics of the flow and conservation laws. Vorticity--based and Lagrangian methods are impractical for the former while Q criterion and $\lambda_2$ are unreliable for the latter. The goal of this study is to propose a novel vortex identification method which can visualize complex turbulent flows while providing individual vortex boundaries and centers that are physically sound so that vortex properties such as loads, circulation, flux can be measured in a way that is flow--agnostic. The process should not rely on complex calculations and should only use variables that are readily available. The proposed alternative shows similar results to the state of the art $\lambda_2$ method for visualizing tripped boundary layer flow and flow over a cylinder while it outperforms it on a lid-driven cavity flow and the flow over a 6:1 prolate spheroid at angle of attack. It is simple, fast to converge and gives a general definition for purpose of quantitative vortex analyses. 

\paragraph*{Theoretical basis}
The momentum equation is written in a translating reference frame following a vortex. Let $\vec{u} = u_i$, $P = P'/\rho$, and $\overline{\overline{\tau}}=\tau_{ij}$ be the instantaneous velocity relative to the center of the vortex, pressure, density and specific shear stress. The incompressible momentum equation can be written in index notation where a repeated index is used to represent summation:
\begin{eqnarray}
\frac{\partial u_i}{\partial t}
+ u_j \frac{\partial u_i}{\partial x_j}
=
-\frac{\partial P}{\partial x_i}
+ \frac{1}{\rho}\frac{\partial \tau_{ij}}{\partial x_j}
\label{eq:momentum}
\end{eqnarray}
The convection term can be written as:
\begin{eqnarray*}
u_j \frac{\partial u_i}{\partial x_j} = \frac{1}{2}\frac{\partial u_j u_j}{\partial x_i} - u_j \omega_k \epsilon_{ijk} e_i
\end{eqnarray*}
where $e_i$ is the unit vector along the $i^{th}$ coordinate and $\epsilon_{ijk}$ is the Levi--Civita symbol. This expression is inserted into equation \ref{eq:momentum}. In addition, using the Taylor's hypothesis of frozen turbulence \citep{taylor1938}, it is assumed that the relative Eulerian velocity inside the vortex is constant, thus $\partial u_i / \partial t = 0$.
\begin{eqnarray*}
\frac{1}{2}\frac{\partial u_j u_j}{\partial x_i} - u_j \omega_k \epsilon_{ijk} e_i
=
-\frac{\partial P }{\partial x_i}
+ \frac{\partial  \tau_{ij}}{\partial x_j}
\end{eqnarray*}

\begin{eqnarray*}
u_j \omega_k \epsilon_{ijk} e_i
=
\frac{\partial}{\partial x_i} (P + \frac{1}{2} u_i u_i)
- \frac{\partial \tau_{ij}}{\partial x_j} 
\end{eqnarray*}
The stagnation pressure is written as: $P_s = P + \frac{1}{2} u_i u_i$, yielding:
\begin{eqnarray*}
u_j \omega_k \epsilon_{ijk} e_i
=
\frac{\partial P_s}{\partial x_i}
- \frac{\partial \tau_{ij}}{\partial x_j} 
\end{eqnarray*}
This yields Crocco's theorem for viscous flows in vector notation:
\begin{subequations}
\begin{eqnarray}
\vec{u} \times \vec{\omega} = \nabla P_s + \nabla \cdot \tau
\label{eq:crocco_viscous}
\end{eqnarray}
$\vec{u} \times \vec{\omega}$  is referred to as the Lamb vector where "$\times$" is the cross product.
If the Reynolds number is sufficiently high, the viscous term is negligible and the equation becomes:
\begin{eqnarray}
\vec{u} \times \vec{\omega} \approx \nabla P_s
\label{eq:crocco_inviscid}
\end{eqnarray}
Note that equation \ref{eq:crocco_inviscid} is instantaneous and allows visualization of turbulent flows assuming steady flow inside the eddies in accordance with Taylor's hypothesis. This hypothesis may not be valid for certain flows, for example, large coherent recirculation vortices. In this case, a time--averaged variant can be used:
\begin{eqnarray}
\langle \vec{u} \times \vec{\omega}\rangle \approx \nabla \langle P_s \rangle
\label{eq:crocco_inviscid_averaged}
\end{eqnarray}
\end{subequations}
Where the bracket $\langle \cdot \rangle$ indicates time averaging.
\begin{center}
\begin{figure}
\begin{tikzpicture}
\draw[even odd rule,inner color=gray,outer color=white, very thick, ->] (9.15,-0.95) circle (1.5);
\filldraw[black] (9.12,-1) circle (2pt) node[anchor=north west]{Vortex center};
\draw[black, very thick, ->] (10.45,-0.2) arc (30:250:1.5)node[anchor=north west]{$P_s = $ constant};
\draw[gray, thick, ->] (10.1,-0.7) arc (15:260:1);
\draw[gray, thick, ->] (9.6,-1) arc (0:280:0.5);
\draw[ultra thick, ->] (7.75,0.1) arc (-180:140:0.3) node[anchor=south east]{$\vec{\omega}$};
\draw[red, ultra thick, ->] (8.05,0.1) -- (7,-1) node[anchor=north west]{$\vec{u}$};
\draw[blue, ultra thick, ->] (8.05,0.1) -- (8.75,-0.6) node[anchor=south west]{$\vec{u} \times \vec{\omega}$};
\end{tikzpicture}
\caption{Schematic of Crocco's theorem applied to a vortex}
\label{fig: condition 1}
\end{figure}
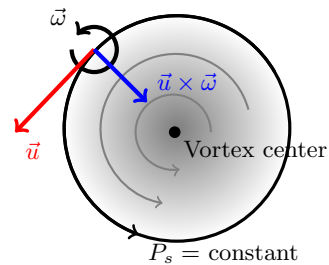
\end{center}

Figure \ref{fig: condition 1} illustrates the terms of equation \ref{eq:crocco_inviscid}. The velocity vector follows the streamline by definition and $\nabla P_s$ is orthogonal to the streamline since $\vec{u} \times \vec{\omega} = \nabla P_s$. This means that streamlines are tangential to the isosurfaces of $P_s$. Since $\vec{u} \times \vec{\omega}$ is also orthogonal to $\vec{\omega}$, the vorticity vectors are also tangential to the isosurfaces of $P_s$. This result is discussed by \citet{truesdell1954} in restrictive `complex-lamellar' cases where $\vec{u} \cdot \vec{\omega} = 0$ and $\|\vec{\omega}\| = 0$. The $\vec{u} \times \vec{\omega}$ isosurfaces are referred to as Lamb surfaces. These surfaces have the interesting property of containing both the streamlines and the vortex lines.
An intuitive way of defining a vortex is to consider the interior of closed streamlines, however, it is difficult to assess whether streamlines are closed in practice. They also depend on the coordinate frame and may not close at all if the vortex is three--dimensional. This issue can be resolved by observing that a closed streamline will be contained inside a tube of constant stagnation pressure since the streamlines are tangential to the Lamb surfaces. Thus a convenient way to define a vortex is to consider a tubular Lamb surface, which has several advantages over more commonly used vortex identification methods, in the inviscid limit:
\begin{itemize}
    \item The Lamb tube is a material boundary which ensures conservation of circulation along its axis; 
    \item Because the streamlines are tangent to the Lamb surfaces, the boundary of the vortex is consistent with a Lagrangian definition while relying on Eulerian metrics, which are simpler to evaluate;
    \item The net flux of momentum and vorticity across the boundary is zero hence the loads integrated over the volume of a vortex core is zero. The loads of the vortex over a nearby wall originate from the intersection of Lamb surfaces and the wall.
\end{itemize}

\begin{center}
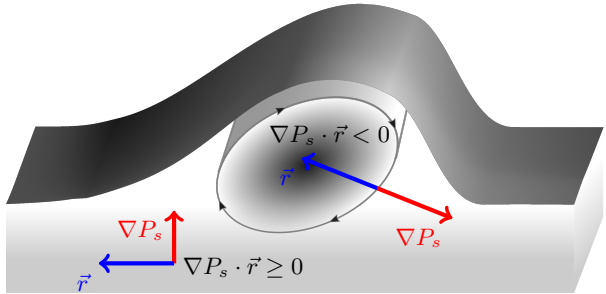
\begin{figure}
\begin{tikzpicture}
\node (drawing) at (8.3,-0.5) {\includegraphics[ width=0.45\textwidth]{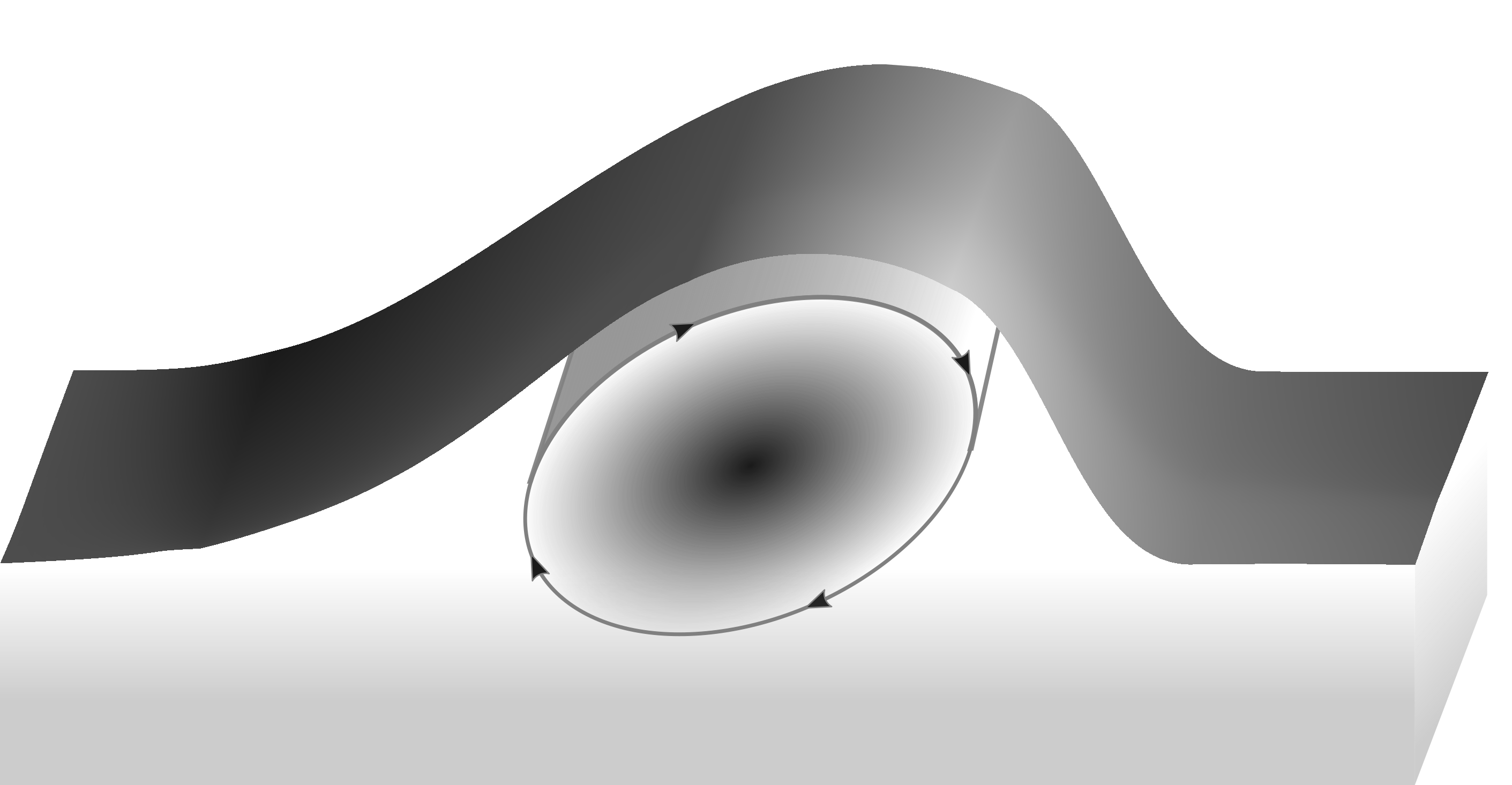}};
\draw[red, ultra thick, ->] (9.2,-1.2) -- (10.2,-1.6) node[anchor=north east]{$\nabla P_s$};
\draw[blue, ultra thick, ->] (9.2,-1.2) -- (8.2,-0.8) node[anchor= north east]{$\vec{r}$};
\filldraw[black] (7.65,-0.25) node[anchor=north west]{$\nabla P_s \cdot \vec{r} < 0$};
\draw[red, ultra thick, ->] (6.5,-2.2) -- (6.5,-1.5) node[anchor=north east]{$\nabla P_s$};
\draw[blue, ultra thick, ->] (6.5,-2.2) -- (5.5,-2.2) node[anchor= north east]{$\vec{r}$};
\filldraw[black] (6.5,-2.0) node[anchor=north west]{$\nabla P_s \cdot \vec{r} \geq 0$};
\end{tikzpicture}

\caption{Schematic illustrating the $\nabla P_s \cdot \vec{r}$ condition for a vortex in a shear layer.}
\label{fig:condition 2}
\end{figure}
\end{center}

The right-hand side of equation \ref{eq:crocco_inviscid_averaged} can be expanded using a Reynolds decomposition where $u_i'$ is the velocity perturbation:
\begin{eqnarray*}
    \nabla \langle P_s \rangle = \nabla (\langle P \rangle + \frac{1}{2} \langle u_i \rangle \langle u_i \rangle + \frac{1}{2} \langle u_i'u_i' \rangle)
\end{eqnarray*}
This form shows that equation \ref{eq:crocco_inviscid_averaged} applied to a vortex is a statement of energy conservation, stored in the form of potential energy $P$, mean kinetic energy $\frac{1}{2} \langle u_i \rangle \langle u_i \rangle$ and turbulent kinetic energy $\frac{1}{2} \langle u_i'u_i' \rangle$. Consider the curvature vector $\vec{r}$ (which can be obtained by normalization of $\partial \vec{u}/\partial s = \partial u_i/\partial x_j \cdot u_j /\|\vec{u}\|$, where $s$ is the streamwise coordinate), pointing from in the direction of the vortex center thus co-linear with $\nabla P_s$. Moving in the direction of $\vec{r}$ from the periphery to the center, the total energy decreases as the enclosed region contains less integrated vorticity thus $\nabla P_s \cdot \vec{r} < 0$. This is illustrated in figure \ref{fig:condition 2}, showing that within the vortex, the gradient is not only aligned with the direction of curvature, but $P_s$ also decreases toward the center. Outside the vortex however, $P_s$ either decreases outward or is constant with the radius. $\vec{r}$ on the other hand, always points toward the center of rotation by definition. Thus the sign of $\nabla P_s \cdot \vec{r}$ can inform the region dominated by the vortex. 
The following criteria are proposed to identify the region of a vortex:
\begin{itemize}
    \item Vortex core: $P_s \leq P_s^{0}$ where $P_s^{0}$ is the value of a closed isoline;
    \item Outer vortex region: $\nabla P_s \cdot \vec{r} < 0$, with $\|\nabla P_s \| \neq 0$. $\vec{r}$ can be calculated by normalization of the vector $\partial u_i / \partial x_j \cdot u_j$.
\end{itemize}
The first condition ensures that the vortex boundary is a material surface, more specifically a vortex tube that conserves circulation along its axis and has zero momentum deficit hence zero load. The second condition identifies regions where pressure gradients are aligned with the local streamline curvature. Lower threshold than zero may be used, similar to common methods such as $\lambda_2$ and Q criterion.
The stagnation pressure field also provides another desirable feature of vortex identification methods, which is to locate the center. At a vortex center, the velocity and the vorticity vectors are co-linear, aligned with the axis of the vortex by axisymmetric constraint, thus the normalized helicity density is unity and normal of the Lamb vector $\|\vec{u} \times \vec{\omega}\|$ is close to zero. The center of the vortex can then be found as a local minimum of stagnation pressure and its gradient.

\paragraph*{Approximation of Lamb surfaces for experimental purposes}
The pressure field in the fluid may not be readily experimentally, however, in certain conditions, it is possible to approximate Lamb surfaces with surfaces of constant total kinetic energy $K = \frac{1}{2} u_i u_i$. These iso-surfaces will approach Lamb surfaces when $\partial P / \partial x_j \ll K$, which is a similar approximation as the one considered by kinematic methods such as $\lambda_2$. In these condition, vortices can be visualized by $\nabla K \cdot \vec{r} \leq 0$ surfaces.

\paragraph*{Connection with vortex loads}
Defining vortices based on isolines offers a unique perspective on the vortex loads since the stagnation pressure appears in the momentum equation. Integrating equation \ref{eq:crocco_inviscid_averaged} over a control volume $V$, bounded by a surface $S$ and normal $\vec{n}$ gives:
\begin{eqnarray*}
\int_V \langle \vec{u} \times \vec{\omega}\rangle dV = \int_S \langle P_s \rangle \vec{n} dS
\end{eqnarray*}
which can be written as:
\begin{eqnarray}
F_{\omega} = \int_S \langle P_s \rangle \vec{n} dS
\label{eq:vortex_force_ps}
\end{eqnarray}
where:
\begin{eqnarray}
F_{\omega} = \int_V \langle \vec{u} \times \vec{\omega}\rangle dV
\label{eq:vortex_force_uw}
\end{eqnarray}
represents the net momentum from vorticity in the control volume \cite{howe1995}, \cite{saffman1995}, which equates the divergence of the stagnation pressure. If the boundary of a vortex is defined as a closed isoline of stagnation pressure of constant value $P_s^{iso}$, the equation becomes:
\begin{eqnarray*}
F_{\omega} = \int_S \langle P_s^{iso} \rangle \vec{n} dS = \langle P_s^{iso} \rangle \int_S \vec{n} dS = 0
\end{eqnarray*}
This is consistent with the result that the net momentum from an isolated vortex is zero. If the vortex is near a wall, the loads come from the intersection of the isolines of stagnation pressure and such wall. Indeed, for that case equation \ref{eq:vortex_force_ps} becomes:
\begin{eqnarray}
F_{\omega}
= \langle P_s^{iso} \rangle \int_{S_{fluid}} \vec{n} dS + \int_{S_{wall}} 
\langle P_s^{wall} \rangle \vec{n} dS
\label{eq:vortex_force_integral}
\end{eqnarray}
Where $S = S_{fluid} \cup S_{wall}$, the fluid and wall surfaces of the control volume respectively.
Consider a spanwise vortex above a horizontal wall located at $y = 0$ and apply equation \ref{eq:vortex_force_integral} in a semi-infinite control volume in $x$ and $y$. Note that the equations are derived in two-dimensions for ease of reading although extension to three-dimensions is trivial.
\begin{eqnarray}
F_{\omega} \cdot \vec{n}_y = \langle P_s^{iso} \rangle \int_{S_{fluid}} \vec{n} \cdot \vec{n}_y dS + \int_{S_{wall}} \langle P_s^{wall} \rangle dS \nonumber\\
= -P_s^{\infty} \int_{x_{-\infty}}^{x =+\infty} dx + \int_{S_{wall}} \langle P_s^{wall} \rangle dx \nonumber\\
= \int_{x = -\infty}^{x = +\infty} \langle P^{wall} - (P^{\infty} + \frac{1}{2} U_{\infty}^2)\rangle dx
\label{eq:vortex_force_example1}
\end{eqnarray}
Where $P_{wall} = P'_{wall}/\rho$ is the static pressure at the wall, $P^{\infty}$ and $U^{\infty}$ are the freestream pressure and velocity respectively. Equation \ref{eq:vortex_force_example1} correctly recovers the expected expression of force. In addition, if the domain is size $l_x$ in the $x$ direction with $l_x \to 0$:
\begin{eqnarray}
lim_{x \to 0} F_{\omega} \cdot \vec{n}_y
= \int_{y = 0}^{y = +\infty} \langle P^{wall} - (P^{\infty} + \frac{1}{2} U_{\infty}^2)\rangle  \nonumber\\
= \int_{y = 0}^{y = +\infty} \frac{1}{2} U_{\infty}^2 (c_p - 1)
\label{eq:vortex_force_example2}
\end{eqnarray}
Where $c_p = 2 (\langle P^{wall} \rangle - P^{\infty}) / U_{\infty}^2$ is the wall pressure coefficient. Combining equations \ref{eq:vortex_force_uw} and \ref{eq:vortex_force_example2} yields:
\begin{eqnarray}
\int_{y = 0}^{y = +\infty} \langle \vec{u} \times \vec{\omega} \rangle dy = \frac{1}{2} U_{\infty}^2 (c_p - 1)
\label{eq:cp_equation}
\end{eqnarray}
Equation \ref{eq:cp_equation} can be used to identify the contribution of specific area of the flow to the wall pressure.

\paragraph*{Procedure}
In summary, the following procedure is used to analyze vortical flows.

Calculate stagnation pressure $P_s = P + \frac{1}{2} \vec{u} \cdot \vec{u}$ for unsteady flows or $\langle P_s \rangle = \langle P \rangle + \frac{1}{2} \langle u_i \rangle \langle u_i \rangle + \frac{1}{2} \langle u_i'u_i' \rangle$ for statistically steady flows. Visualization of this variable yields: regions where the flow is rotational (boundary layers, recirculations...) with values of $P_s$ that are smaller than the free-stream value; vortex centers as local minima of stagnation pressure; vortex core as region contained in closed isolines of stagnation pressure; vortex boundary as largest closed isoline of stagnation pressure. This boundary can be used for the purpose of integral calculation on the vortex.

This approach is simple and yields reliable results for the analysis of a 2D slice of isolated vortices. The following procedure is followed for the purpose of visualizing 3D isosurface in general turbulent flows:
\begin{itemize}
    \item Calculate the gradient of stagnation pressure as:
    \[
        \nabla P_s = \frac{\partial P_s}{\partial x_j} = \frac{\partial P}{\partial x_j} + u_i \frac{\partial u_i}{\partial x_j}  
    \]
    \item Calculate the vector $\vec{r}$, which points toward the center of rotation of the streamline as:
    \[
      \vec{r} =  \frac{u_j \frac{\partial u_i}{\partial x_j}}{\sqrt{u_k u_k}}
    \]
    \item Calculate $\nabla P_s \cdot \vec{r}$ and visualize its negative iso--surfaces.
\end{itemize}

\paragraph*{Analytical vortex: Rankine model}
The Rankine vortex is a simplified two-dimensional model where the fluid moves in solid body rotation within a radius $r_0$ and where the flow is potential at a distance greater than $r_0$:
\begin{eqnarray*}
\begin{cases}
u_r = u_z = 0 \\
u_{\theta} = u_0 \frac{r}{r_0}, r \leq r_0 \\
u_{\theta} = u_0 \frac{r_0}{r}, r \geq r_0
\end{cases}
\end{eqnarray*}
The momentum equation in cylindrical coordinate simplifies to
\begin{eqnarray*}
\frac{\partial P}{\partial r} = \frac{u_{\theta}^2}{r}
\end{eqnarray*}
Thus the $r$ derivative of the stagnation pressure is:
\begin{eqnarray*}
\frac{\partial P_s}{\partial r} = \frac{\partial P}{\partial r} + u_{\theta} \frac{\partial u_{\theta}}{\partial r}
\end{eqnarray*}
\begin{eqnarray*}
=
\begin{cases}
\frac{u_0^2 r}{r_0^2} + \frac{u_0^2 r}{r_0^2} = 2\frac{u_0^2 r}{r_0^2}, r < r_0 \\
\frac{u_0^2 r_0^2}{r^3} - \frac{u_0^2 r_0^2}{r^3} = 0, r > r_0
\end{cases}
\end{eqnarray*}
If the vortex is taken at the boundary where $\nabla P_s \cdot \vec{r} = 0$, in this special case $\partial P_s / \partial r = 0$, the radius of vortex is perfectly recovered as $r_0$, thus the method is thus consistent for a Rankine vortex.
In the next paragraphs, the vortex identification method is tested on several numerical cases.

\paragraph*{Numerical approach}
The incompressible, spatially filtered Navier-Stokes equation are solved in a Large-Eddy Simulation formulation:
\begin{eqnarray}
\frac{\partial \overline{u}_i}{\partial t} + \frac{\partial}{\partial x_j}(\overline{u}_i \overline{u}_j) & = &
-\frac{\partial \overline{P}}{\partial x_i}
+ \nu \frac{\partial^2 \overline{u}_i}{\partial x_j \partial x_j}
- \frac{\partial \overline{\tau}_{ij}}{\partial x_j}, \\
\frac{\partial \overline{u}_i}{\partial x_i} & = & 0,
\end{eqnarray}
The Sub-Grid Stress (SGS) $\overline{\tau}_{ij} = \overline{u_iu_j}-\overline{u}_i\overline{u}_j$ is modeled with the dynamic Smagorinsky model \citep{germano1991,lilly1992}.
A finite volume, second order centered spatial discretization is used where the filtered velocity components and pressure are stored at the cell-centroids while the face-normal velocities are estimated at the face centers. The equations are marched in time with a second order Crank-Nicolson scheme.
A diverse range of complex flow have been studied using the formulation, such as propeller in crashback \citep{verma2012, kroll2022} and flow over hulls \citep{kumar2017,morse2021}. The kinetic-energy conservation property of the method \citep{mahesh2004} makes it suitable for high Reynolds number flow such as the one presented in this paper.
\citet{horne2019b} extended the method for overlapping (overset) grids and six-degree of freedom. Although all the geometries presented in the current study are static, the overset formulation is used to limit the computational cost while increasing the resolution in regions of high shear.

\begin{figure}[ht]
\begin{center}
    \includegraphics[trim={0 0 0 0}, clip, width = 80 mm]{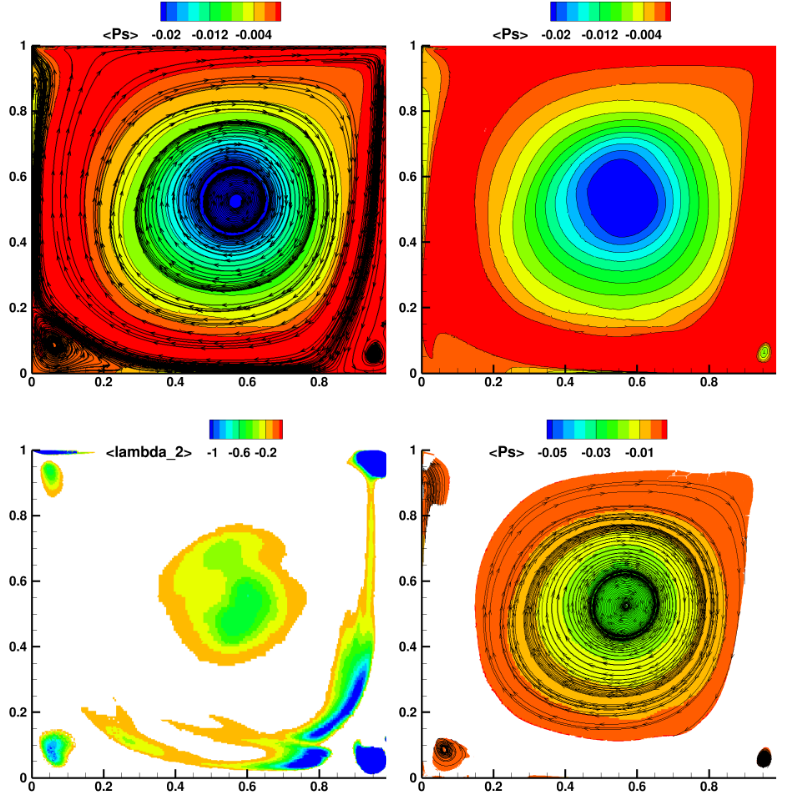}
    \put(-215,225){$a)$}
    \put(-102,225){$b)$}
    \put(-215,110){$c)$}
    \put(-102,110){$d)$}
    \caption{Averaged velocity streamlines (a); mean stagnation pressure isolines (b); $\lambda_2$ masked at $\lambda_2 < -0.1$ (c); and mean stagnation pressure masked by $\nabla P_s \cdot \vec{r} < 0$ (d) in the cavity at $Re = 10000$.}
    \label{fig:cavity lambda2 and Ps}
\end{center}  
\end{figure}
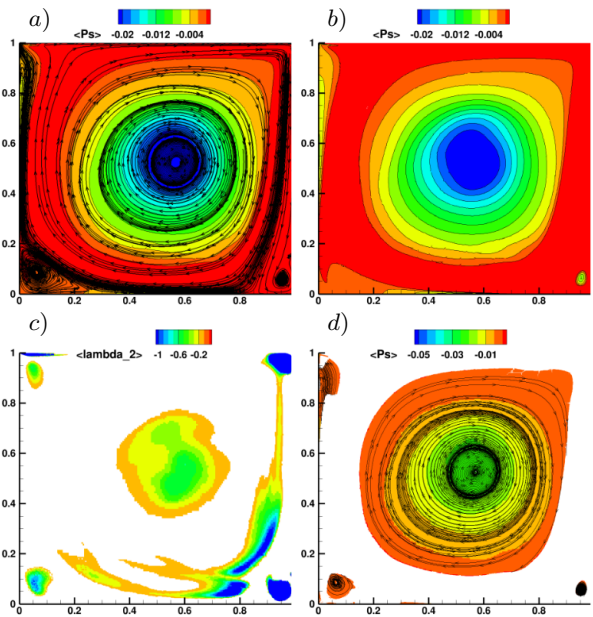

\paragraph*{Case I - Cavity flow}
\label{sec: cavity flow}
A cavity flow is a canonical problem which produces a steady vortex. The chosen Reynolds number is 10000, for a cubic cavity of unit length in all directions. The top boundary has unit velocity $U_0$, no-slip condition are applied on the bottom and sides walls with a periodic boundary on the span. The flow is developed until the loads plateau, which happens at around 40 units. 
Figure \ref{fig:cavity lambda2 and Ps} shows the averaged velocity streamlines, the mean stagnation pressure isolines, $\lambda_2$ and the mean stagnation pressure blanked by $\nabla P_s \cdot \vec{r}$. Even for this moderate Reynolds number, the isolines of stagnation pressure are very well aligned with the streamlines. The deficit of stagnation pressure correctly identify the location of the primary vortex and the three secondary vortices. In contrast, $\lambda_2$ correctly identifies the primary as well as the secondary vortices, however, the shape of the vortices deviates from the streamlines. In addition, it has several false positives, specifically on the top corner and below the primary vortex. These artifacts occur where the streamlines have significant curvature though no vortex is present. The $\nabla P_s \cdot \vec{r}$ criterion correctly identifies all the vortical structures without false positives; the iso-surfaces of stagnation pressure are well aligned with the streamlines.
As previously mentioned, the method can also be used to estimate loads. Applying the equation \ref{eq:crocco_viscous} along the vertical direction and integrating over the cavity yields:
\begin{eqnarray}
\int_V \langle \vec{u} \times \vec{\omega} \rangle \cdot \vec{n}_y dV = - \int_{S_{top}} \langle P_s \rangle dS \nonumber\\
+ \frac{1}{2} U_0^2 (\int_{S_{left}} c_f dS - \int_{S_{right}} c_f dS + \int_{S_{bottom}} c_p dS) 
\label{eq: cavity equation}
\end{eqnarray}
where $V$ is the volume of the cavity, $S_{left}$, $S_{right}$, $S_{bottom}$ and $S_{top}$ are the surfaces of the back and front walls, $c_f$ is the skin friction coefficient, $c_p$ is the wall pressure coefficient, $P_s$ is the stagnation pressure.

\begin{table}[b]
\caption{\label{tab:cavity}}
\begin{ruledtabular}
\begin{tabular}{cccc}
\textrm{$\langle\vec{u}\times\vec{\omega}\rangle_y$}&
\textrm{$F_y$ top}&
\textrm{$F_y$ sides}&
\textrm{$F_y$ bottom}\\
\colrule
$0.5071$ & $0.5056$ & $-4.077e^{-3}$ & $-2.064e^{-3}$ \\
\end{tabular}
\end{ruledtabular}
\end{table}

\begin{figure}[ht]
\begin{center}
    \includegraphics[trim={40 40 40 40}, clip, width = 60 mm]{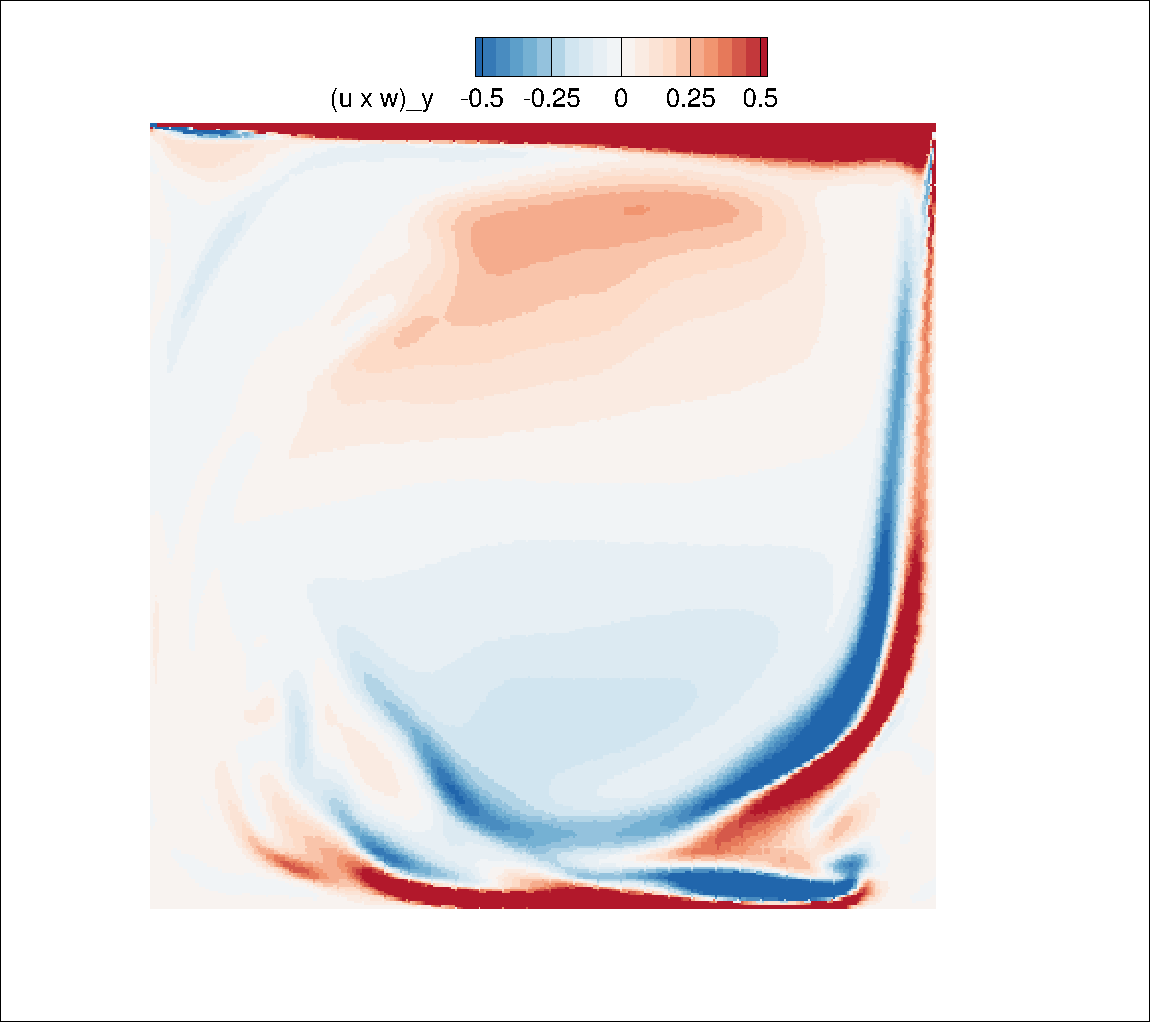}
    \caption{$\langle \vec{u} \times \vec{\omega} \rangle_y$ in the cavity in the steady state}
    \label{fig:cavity uxw_y}
\end{center}  
\end{figure}

Table \ref{tab:cavity} shows the value of each of the terms of equation \ref{eq: cavity equation}. Even for this moderate Reynolds number, the value of the viscous term is negligible (two orders of magnitude) compared to the pressure terms and the vorticity term. This justifies the inviscid form of Crocco's theorem for vortical flows with higher Reynolds number based on vortex diameter. This also suggests that the Lamb vector can be used as a way to measure the load contribution of a fluid parcel on the wall. Figure \ref{fig:cavity uxw_y} shows its local value, representative of the contribution of the vortical force to the wall: a negative value is indicative of an increased pressure contribution on the bottom wall while a positive value indicates suction. Most of the loads come from the region on the edge of the primary vortex and are suctions, except at the bottom where the separated sheet impinges on the bottom wall and contributes to an increased pressure on that boundary.

\begin{figure}[ht]
\begin{center}
    \includegraphics[trim={10 10 10 10}, clip, width = 60 mm]{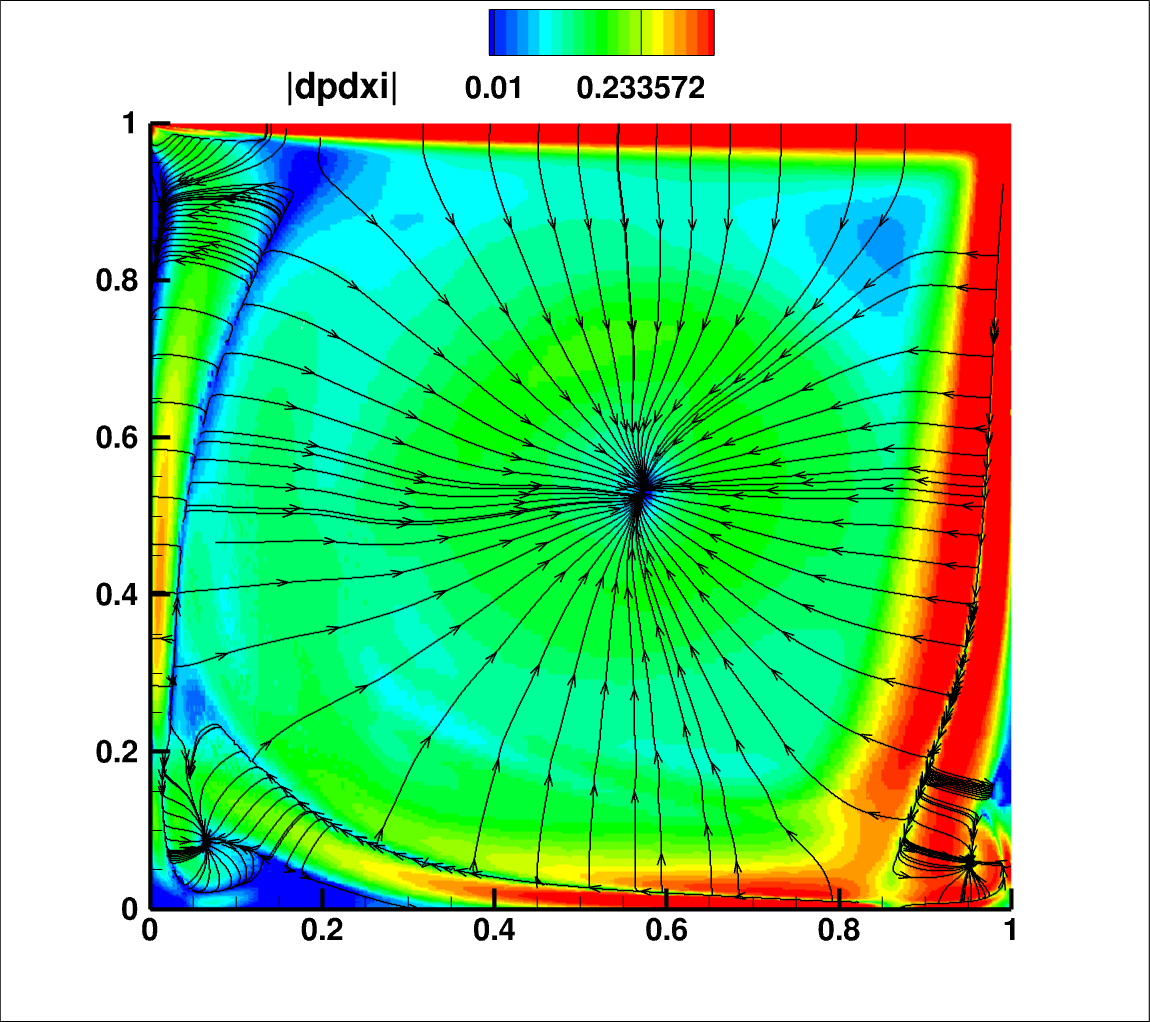}
    \caption{Color: $|\langle\nabla P_s\rangle|$ in the cavity. The lines are tangential to the direction of $\langle\nabla P_s\rangle$}
    \label{fig:cavity gradient}
\end{center}  
\end{figure}

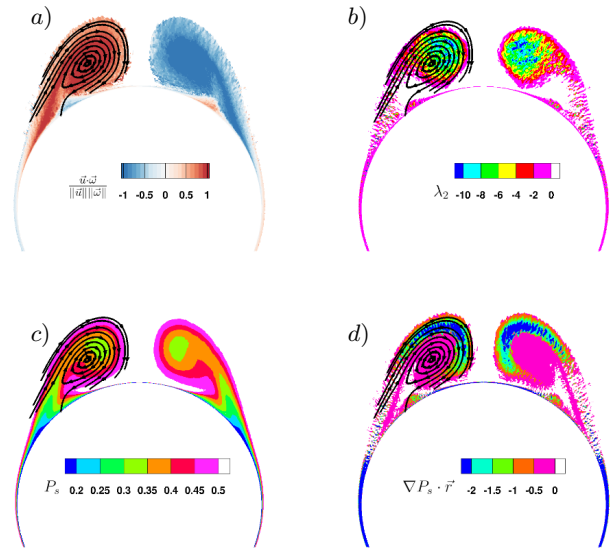
\begin{figure}[ht]
\begin{center}
    \includegraphics[width = 80 mm]{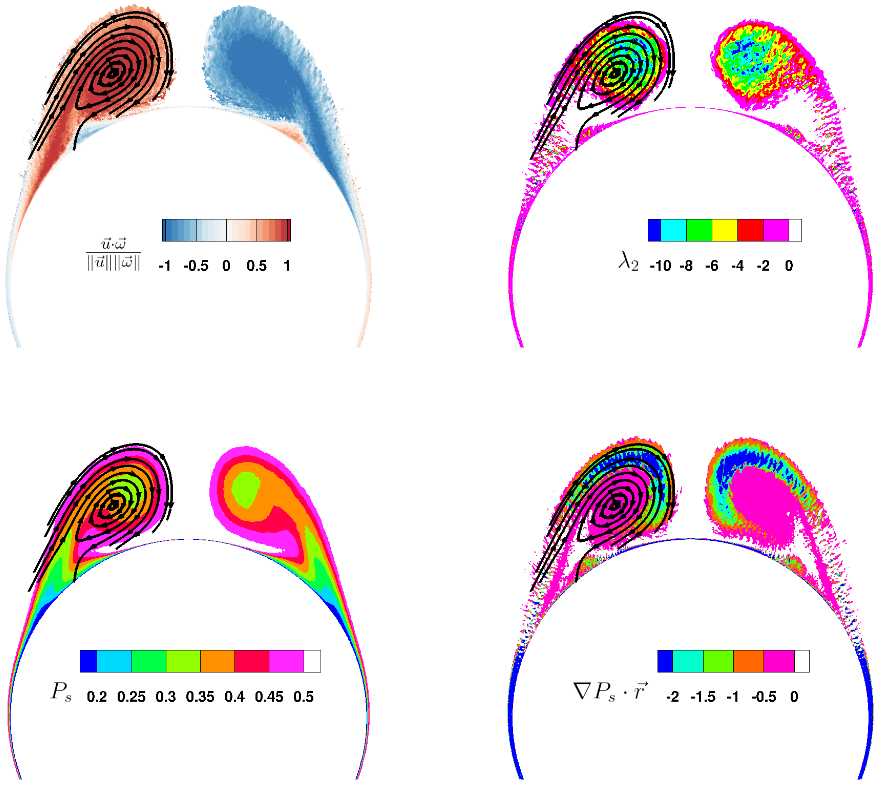}
    \put(-220,200){$a)$}
    \put(-100,200){$b)$}
    \put(-220,80){$c)$}
    \put(-100,80){$d)$}
    \caption{Secondary streamlines and normalized helicity density (a); $\lambda_2$ (b); time-averaged stagnation pressure (b); $\langle \nabla P_s \cdot \vec{r}\rangle$ (d) in a transverse plane at $x/L = 0.772$, in the prolate spheroid flow at $20^\circ$ incidence}
    \label{fig:spheroid_transverse_20deg}
\end{center}  
\end{figure}

Figure \ref{fig:cavity gradient} shows the direction and magnitude of $\langle \nabla P_s \rangle$, demonstrating equation \ref{eq:crocco_inviscid} and figures \ref{fig: condition 1} and \ref{fig:condition 2} for a simulated flow. Inside the vortices, the gradient points toward the center of rotation and are orthogonal to the streamlines seen in figure \ref{fig:cavity lambda2 and Ps}. The lines converge toward localized minima of stagnation pressure $\langle P_s \rangle$ which constitute the centers of the vortices.
\paragraph*{Case II - 6:1 prolate spheroid}
\label{sec: prolate spheroid}
The 6:1 prolate spheroid is a canonical geometry that is commonly used to study smooth three-dimensional separation. At angle of attack, the boundary layer separates and forms a coherent, attached, counter-rotating vortex pair. A  $20^\circ$ incidence of the flow is chosen with a Reynolds number $Re_L = 4.2M$ where $L$ is the length of the geometry. Numerical details are available in \citet{plasseraud2023}, from which these results are taken. The flow is computed using wall-resolved large-eddy simulation on a 600M control volume overset grid tripped at $x/L = 0.2$.  The statistics are averaged for one spheroid flow-through of time. In these conditions, the boundary layer separates along two separate longitudinal lines and two counter-rotating vortex pairs are formed, a primary pair and a smaller secondary pair underneath the primary separation sheet. A certain number of features make this flow particularly challenging to simulate and for vortex identification methods \citep{plasseraud2023}:
\begin{itemize}
    \item The flow is characterized by several smooth 3D separations leading to several counter-rotating vortex pairs on the lee of the obstacle. Smooth separations are hard to resolve because they depend on the state of turbulence of the boundary layer; 3D separations are challenging to locate because they cannot be identified by the cancellation of the near-wall velocity or skin friction;
    \item The vortex pairs are attached, which may be problematic for vortex identification since it is hard to isolate where the separated sheet ends and where the vortex begins. Isolating the separated sheet may be desirable for the purpose of analysis;
    \item The vortices are strongly three-dimensional, which means that there is a net flux of momentum along their axis and that the streamlines are three-dimensional curves, complicating the vortex identification especially for Lagrangian methods;
    \item The multiple vortices and image vortices interact with each other to modify the velocity field. This implies that the center of each vortex has non-zero velocity magnitude;
    \item The surface of the spheroid is curved thus identification methods based on streamline curvature alone (i.e. Q and $\lambda_2$ method) may interpret the attached flow as a vortex;
    \item The separation sheet and the primary vortex pairs have normalized helicity magnitude that are close to unity (Beltrami flow), where the velocity vector and the vorticity vectors are almost co-linear even though they have non-zero norm. This may challenge the proposed method that relies on Lamb surfaces, which could become undefined in Beltrami flows.
\end{itemize}  
Figure \ref{fig:spheroid_transverse_20deg} (a) shows secondary streamlines and normalized helicity density on the lee of the prolate spheroid at $x/L = 0.772$, where $x$ is the axial coordinate. The streamlines display the primary vortex pair as well as the separation sheet. The helicity density shows the primary vortex pair, the separated sheet as well as the small counter-rotating secondary vortex pair close to the wall. The helicity is very close to one in both the separation sheet and in the center part of the primary recirculation region; while its absolute value is between $0.5$ and $1$ in the outer part of that area.\ref{fig:spheroid_transverse_20deg}(b): $\langle\lambda_2\rangle$ captures the general area around the vortex however the boundary does not align well with the streamlines and it does not represent the elliptical shape seen on the streamlines and in helicity. Furthermore, the center of the primary vortices cannot be reliably located as the minimum is too noisy. Similar values of $\langle \lambda_2 \rangle$ are seen between the attached boundary layer, the separation sheet and the entire secondary vortices, making it hard to discriminate between these three features. This lack of contrast comes from the difficulty the $\lambda_2$ method has to eliminate attached curved streamlines from detached one as discussed above.
\ref{fig:spheroid_transverse_20deg}(c) shows the time-averaged stagnation pressure. The core is distinctly visible as a closed circular region of lower pressure. The separation is identifiable as a saddle in this transverse plane. The isolines on the outer part of the recirculation and the separation agree well with the streamlines although the predicted location of the core is offset from the minimum of secondary velocity, similarly to $\lambda_2$. This is due to the induced velocity from the other real and mirror vortices as discussed previously.
\ref{fig:spheroid_transverse_20deg}(d): $\langle \nabla P_s \cdot \vec{r}\rangle$ agrees even better with the secondary streamlines. The separation sheet and the inner part of the vortex where the helicity is unity, are distinctly visible as a negative, close to zero value. The separation sheet is thin and allows to both identify the location of separation and the point of contact with the primary vortex. Surprisingly, another region is visible in the recirculation, which surrounds about half of the inner vortex and that has lower gradient of stagnation pressure. This pattern, along with the observations on the helicity, shows that the fluid from the separating boundary layer is first advected in the center of the vortex rather than at the periphery. The vorticity is then diffused from the core to the outer outer region that appears with a strong gradient of stagnation pressure. In addition, a secondary vortex pair is identifiable close to the wall and distinct from the turbulent boundary layer. The capacity of the proposed method to distinguish between the various layers of the vortex and clearly isolate the separation and small secondary vortices close to the wall is a unique and invaluable property for the study of complex flows.

The following two cases demonstrate the ability of the method to visualize complex unsteady turbulent flows.

\begin{figure*}[ht]
\begin{center}
    \includegraphics[width = 120 mm]{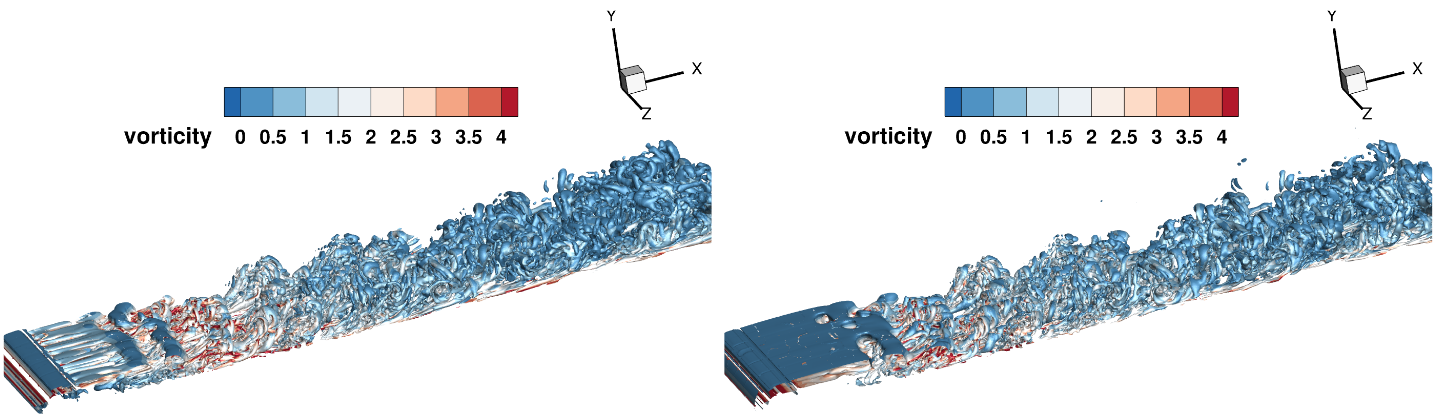}
    \caption{$\lambda_2 = -0.01$ (left) and $\nabla P_s \cdot \vec{r} = -0.1$ isosurfaces (right) colored by instantaneous spanwise vorticity, in the tripped boundary layer case at $Re=1000$.}
    \label{fig:trip}
\end{center}  
\end{figure*}

\paragraph*{Case III - Tripped boundary layer}
\label{sec: trip}
The stagnation pressure method is first used for a flow over a flat plate with zero pressure gradient. The flow is tripped using a horizontal cylindrical wire of unit diameter. The Reynolds number based on trip height is 1000; the inflow is prescribed as a Blasius profile of boundary layer thickness $\delta_{99} = 1.5$. The flow is resolved using wall-resolved large eddy simulation.  More details about the simulation and the results are provided in \citet{plasseraud2022}, from which the results are taken. The challenge of vortex identification methods in this case, is to isolate vorticity inside vortices (vortical vorticity) from vorticity in the boundary layer (shearing) since the wake develops in a boundary layer.
Figure \ref{fig:trip} shows the $\lambda_2 = -0.01$ (left) and $\nabla P_s \cdot \vec{r} = -0.1$ isosurfaces in the region downstream of the trip. Both methods have very similar results, showing the complex development of hairpin vortices in the wake of the trip. The boundary layer upstream of the trip has a Blasius profile, which contains non-vortical vorticity. The proposed method successfully isolate the vortical vorticity from the shearing and does not show any surface from the upstream region. The two methods diverge slightly in the near wake of the trip, where the separation sheet is wavier with the $\lambda_2$ method compared to a flatter surface for the stagnation pressure method. This is due to the former method capturing more rotational motion of the streamwise instabilities while the latter is better at identifying separation with a longer radius of curvature.

\begin{figure*}[ht]
\begin{center}
    \includegraphics[width = 120 mm]{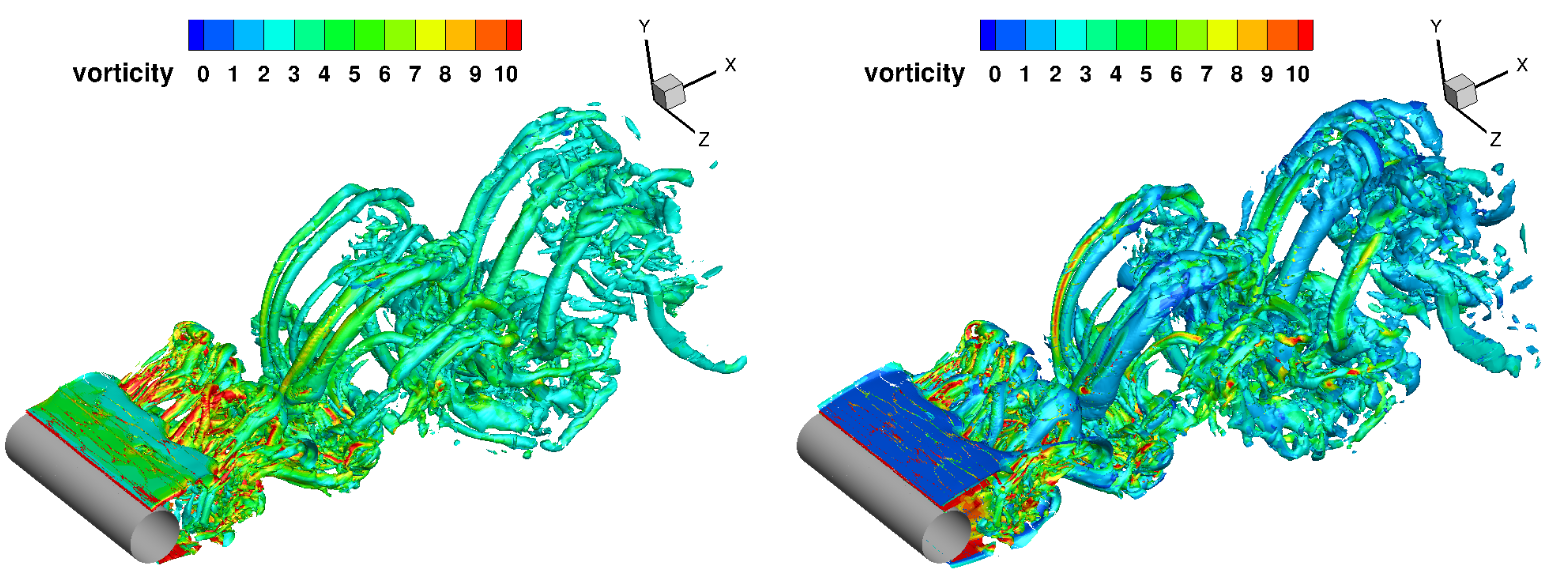}
    \caption{Cylinder flow at $Re = 1000$: $\lambda_2 = -1$ iso-surface (left) and $\nabla P_s \cdot \vec{r} = -0.6$ iso-surface (right), colored by instantaneous vorticity magnitude.}
    \label{fig:cylinder1000}
\end{center}  
\end{figure*}

\paragraph*{Case IV - Cylinder}
\label{sec: cylinder}
The flow around an infinite cylinder for Reynolds number based on diameter of 1000 is calculated. A constant inflow of velocity $U^{\infty}$ is prescribed, with a Neumann boundary as the outflow. The domain is four diameters in the span with 30M control volumes. Figure \ref{fig:cylinder1000} shows the $\lambda_2 = -1$ isosurface (left) and the $\nabla P_s \cdot \vec{r} = -0.6$ isosurface (right), both colored by instantaneous vorticity magnitude. This regime is characterized by alternate periodic shedding of spanwise vortices at measured Strouhal number $St = fD/U^{\infty} \approx 0.196$ (where f is the frequency of shedding, D the diameter of the cylinder) with smaller streamwise, elongated counter-rotating vortices connecting the primary ones. The proposed method is able to capture the unsteady flow in the wake of the cylinder even at low Reynolds number. It also correctly show the primary and secondary structures and give similar results as $\lambda_2$. The stagnation pressure method successfully captures a large range of vortices size, topologies and successfully isolate the streamwise counter-rotating vortex pairs. Some differences are observed between the two methods: the geometry of the vortices are slightly different (visible with different values of vorticity). Similarly to the tripped flow problem (see paragraph `Case III - Tripped boundary layer'), the stagnation pressure method successfully isolate the vortical vorticity from shearing. 

\paragraph*{Conclusion}
A new vortex identification methodolody is proposed for incompressible flows, which uses isosurfaces of stagnation pressure as a vortex boundary. More specifically, a vortex is contained in a region where:
\begin{itemize}
    \item Vortex core: $P_s = P'/\rho + 1/2 u_i u_i \leq P_s^{0}$ where $P_s^{0}$ is the value of a closed isoline;
    \item Outer vortex region: $\nabla P_s \cdot \vec{r} < 0$, with $\|\nabla P_s \| > 0$. $\vec{r}$ is a vector that points towards the local curvature of the flow and can be calculated by normalization of the vector $\partial u_i / \partial x_j \cdot u_j$.
    \item Vortex center: minimum of $P_s$ and $\|\nabla P_s \|$.
\end{itemize}
The method is robust, fast to converge, trivial to use and can be applied to both steady and unsteady flows. The theoretical basis originates from Crocco's theorem which guarantees that in the inviscid limit, streamlines are tangential to stagnation pressure iso-surfaces while the norm of the stagnation pressure gradient is minimal at vortex centers. Despite the assumption of inviscid flow, the method has demonstrated excellent identification abilities even at moderate Reynolds numbers. The method can be used to help estimate vortex loads by using the relation between $\vec{u} \times \vec{\omega}$ and the isoline of stagnation pressure. For experimental purposes where the pressure field is not readily available, vortices can instead be visualized with iso-surfaces of $\nabla \frac{1}{2} u_i u_i \cdot \vec{r} < 0$, which approximates the stagnation pressure criterion when the pressure gradient is negligible compared to the gradient of kinetic energy.

\begin{acknowledgments}
 Mr Ali Fakhreddine, Dr Praveen Kumar, Mr Aditya Madabhushi, Dr Nicholas Morse and Mr Soham Prajapati for the numerous discussions of vortex identification. Computing resources were provided through a United States Department of Defense (DoD)
Frontier project of the High Performance Computing Modernization Program (HPCMP) by the US Army
Engineer Research and Development Center (ERDC) in Vicksburg, Mississippi, on the Onyx supercomputer
of the High Performance Computing Modernization Program.
\end{acknowledgments}

This work is supported by the United States Office of Naval Research (ONR) under ONR Grant N00014-20-1-2717 with Dr. Peter Chang as technical monitor.

\bibliography{apssamp}.
\end{document}